\documentclass[a4paper,11pt]{article}
\usepackage{pos}
\usepackage{natbib}
\usepackage{soul}

\title{Unraveling the Effects of Dense Medium on a Near to Bohm-Limit Acceleration in Kepler's SNR}

\author*[a,b]{Vincenzo Sapienza}
\author[a,b]{Marco Miceli}
\author[b,c]{Oleh Petruk}
\author[d,e,f]{Aya Bamba}
\author[b]{Salvatore  Orlando}
\author[b]{Fabrizio Bocchino}
\author[a,b]{Giovanni Peres}

\affiliation[a]{Dipartimento di Fisica e Chimica E. Segr\`e, Universit\`a degli Studi di Palermo\\ Piazza del Parlamento 1, 90134, Palermo, Italy}

\affiliation[b]{INAF-Osservatorio Astronomico di Palermo\\ Piazza del Parlamento 1, 90134, Palermo, Italy}

\affiliation[c]{Institute for Applied Problems in Mechanics and Mathematics\\ Naukova Street 3-b, 79060 Lviv, Ukraine}

\affiliation[d]{Department of Physics, Graduate School of Science, The University of Tokyo\\ 7-3-1 Hongo, Bunkyo-ku, Tokyo 113-0033, Japan}

\affiliation[e]{Research Center for the Early Universe, School of Science, The University of Tokyo, 7-3-1 Hongo, Bunkyo-ku, Tokyo 113-0033, Japan}

\affiliation[f]{Trans-Scale Quantum Science Institute, The University of Tokyo\\ 7-3-1 Hongo, Bunkyo-ku, Tokyo 113-0033, Japan}

\emailAdd{vincenzo.sapienza@inaf.it}

\abstract{
The maximum energy of electrons accelerated by supernova remnants (SNR) is typically limited by radiative losses.
In this scenario, the synchrotron cooling time scale is equal to the acceleration time scale. 
On the other hand, the low propagation speed of a shock in a dense medium is expected to result in an extended acceleration time scale, thus inducing a decrease in the maximum electron energy for a given SNR age and in the X-ray nonthermal flux.
The young Kepler's SNR shows an enhanced efficiency of the acceleration process, which is close to the Bohm limit in the north of its shell, where the shock is slowed down by a dense circumstellar medium.
Conversely, in the south, where no interaction with a dense medium is evident and the shock speed is high, the acceleration proceeds with a higher Bohm factor.
To investigate this scenario, we studied the temporal evolution of the non-thermal emission, taking advantage of two \textit{Chandra} X-ray observations of Kepler's SNR (performed in 2006 and 2014).
We analyzed the spectra of different filaments both in the north and south of the shell, and measured their proper motion.
We found a region with low shock velocity where we measured a significant decrease in flux from 2006 to 2014.
This could be the first evidence of fading synchrotron emission in Kepler's SNR.
This result suggests that under a certain threshold of shock speed the acceleration process could exit the loss-limited regime.}

\ConferenceLogo{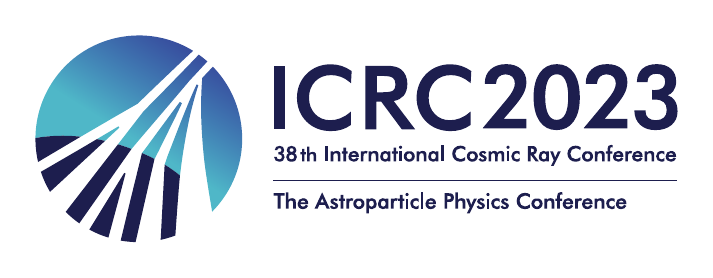}

\FullConference{%
38th International Cosmic Ray Conference (ICRC2023)\\
26 July - 3 August, 2023\\
Nagoya, Japan}

\begin{document}
\maketitle

\section{Introduction}

Supernova Remnants (SNRs) are prominent accelerators of particles, thus they are widely considered the primary foundry of galactic cosmic rays.
The first observational evidence of very high energy (VHE; i.e. $E > 10^{12}$ eV) electrons accelerated at SNR shocks was discovered by \cite{1995Natur.378..255K}, who detected nonthermal X-ray emission stemming from SN 1006.
Indeed, young SNRs typically emit synchrotron radiation in the X-ray band, which can be used as a diagnostic tool to deepen our understanding of the acceleration dynamics.
The study of X-ray synchrotron emission provides information about the electron energy distribution and the mechanism that limits the maximum energy that electrons can reach, such as energy losses by radiation or age limitation.

Kepler's SNR, the aftermath of the explosion of the historical SN 1604, is an interesting object to study the acceleration process.
The remnant, resulting from a Type Ia SN \cite{1999PASJ...51..239K}, is interacting with a dense nitrogen-rich circumstellar medium (CSM) in the north [\citealp{2007ApJ...668L.135R}, \citealp{2021ApJ...915...42K}].
Recent estimates based on proper motion measurements derived a distance $d = 5.1_{-0.7}^{+0.8}$ kpc \citep{2016ApJ...817...36S}.
We then adopt $d=5$ kpc throughout this paper.
Prominent particle acceleration in Kepler's SNR is testified by its energetic non-thermal emission.
The presence of non-thermal X-ray emission in Kepler's SNR was first discovered in its south-eastern region by \cite{2004A&A...414..545C}.
By analyzing a \emph{Suzaku} HXD observation, \cite{2021PASJ...73..302N} provided the first robust detection of hard X-ray emission, in the 15-30 keV band.
A recent study \cite{2021ApJ...907..117T} measured the cutoff photon energy ($\varepsilon_0$) in different regions of several SNRs, Kepler's SNR among them, using a model of synchrotron emission where the maximum energy of the electrons is limited by radiation losses proposed by \cite{2007A&A...465..695Z} (hereafter, loss-limited model).
In this model, $\varepsilon_0$ is related to the shock speed, $v_{sh}$, as 
\begin{equation} 
\label{eq:eovsvsh}
    \varepsilon_0=\frac{1.6}{\eta}\bigg( \frac{v_{sh}}{4000\text{ km s}^{-1}}\bigg)^2 \text{keV},
\end{equation}
where $\eta$, or Bohm factor, is the ratio between the diffusion coefficient and $c\lambda/3$ (where $\lambda$ is the Larmor radius, the minimum value $\eta=1$ corresponding to the Bohm limit) and is strongly related to the turbulence of the magnetic field, which scatter the charged particles.
The spatially resolved analysis by \cite{2021ApJ...907..117T} of Kepler's SNR lacked of the hard part of the spectrum and the $\epsilon_0 - v_{sh}$ plot showed a clear trend only for synchrotron dominated regions, while no correlation could be found for other regions.

In our previous research \cite{2022ApJ...935..152S}, making use of \textit{NuSTAR} and \textit{XMM-Newton} data, we performed a spatially resolved spectral analysis of Kepler's SNR, using the hard part of the X-ray spectrum where the emission is dominated by synchrotron radiation.
The spectra were analyzed by adopting the loss-limited model.
We identified two different regimes of particle acceleration, characterized by different Bohm factors.
In the north, where the shock interacts with a dense circumstellar medium (CSM), we found a more efficient acceleration than in the south, where the shock velocity is higher and there are no signs of shock interaction with dense CSM.
This testifies that the interaction of the shock front with the high density CSM at north is associated with an amplified, turbulent, magnetic field, which enhances the particle acceleration process.
On the other hand, the slow shock in the north \cite{2022ApJ...926...84C} lead to a high acceleration time scale ($\sim$300 yrs), which can result in a decrease of the maximum electron energy by radiation losses.

To unravel this puzzling scenario we studied the evolution of the synchrotron flux in several filaments of Kepler's SNR, making use of the two deepest \textit{Chandra} observations in 2006 and 2014.
This paper is organized as follow: in Section \ref{data} we present the datasets and the data reduction process, whereas in Section \ref{res} we show the results retrieved so far. Preliminary discussions and conclusion are drawn in Section \ref{con}.

\section{Observations and Data Reduction} \label{data}
For our analysis we made use of different \textit{Chandra} observations mainly focused on two years: 2006 and 2014. These are summarized in Table \ref{tab:obs}.
\begin{table}[t!]
    \centering
    \caption{\textit{Chandra} observations table.}
    \begin{tabular}{ccccc}
    \hline\hline
    Obs ID & Exp. Time (ks) & R. A.                  & Dec.                      & Start Date \\
    \hline
    6714   & 157.8          & 17$^h$ 30$^m$ 42.0$^s$ & -21\textdegree 29' 00.0'' & 27/04/2006 \\
    6715   & 159.1          & 17$^h$ 30$^m$ 41.2$^s$ & -21\textdegree 29' 31.4'' & 03/08/2006 \\
    6716   & 158.0          & 17$^h$ 30$^m$ 42.0$^s$ & -21\textdegree 29' 00.0'' & 05/05/2006 \\
    6717   & 106.8          & 17$^h$ 30$^m$ 41.2$^s$ & -21\textdegree 29' 31.4'' & 13/07/2006 \\
    6718   & 107.8          & 17$^h$ 30$^m$ 41.2$^s$ & -21\textdegree 29' 31.4'' & 21/07/2006 \\
    7366   & 51.5           & 17$^h$ 30$^m$ 41.2$^s$ & -21\textdegree 29' 31.4'' & 16/07/2006 \\
    16004  & 102.7          & 17$^h$ 30$^m$ 41.2$^s$ & -21\textdegree 29' 31.4'' & 13/05/2014 \\
    16614  & 36.4           & 17$^h$ 30$^m$ 41.2$^s$ & -21\textdegree 29' 31.4'' & 16/05/2014 \\
    \hline
    \end{tabular}
    \label{tab:obs}
\end{table}
The data were reprocessed with the CIAO v4.13 software using CALDB 4.9.4.
We reprocessed the data by using the \texttt{chandra\_repro} task. 
We mosaicked flux images for each year by using the \texttt{merge\_obs} task.
We followed the same astrometric alignment procedure described by \cite{2022ApJ...926...84C}, using the same point sources (magenta ellipses in Figure \ref{fig:reg}), to conduct a proper motion measurement for our purposes.
For the proper motion measurement only, we use the deep 2006 observation (Obs. ID: 6715) as the relative reference to which we aligned the 2014 deepest observation (Obs. ID: 16004).
To extract spectra from all the observation, we used \texttt{specextract} CIAO command.
We then combined the spectra from the same epochs by using the \texttt{combine\_spectra} CIAO command.
The spectra were binned using the optimal binning algorithm \cite{2016A&A...587A.151K}.
The spectral analysis was conducted on XSPEC v. 12.11.1 \cite{1996ASPC..101...17A} and for the fitting procedure the Cash statistic (C-stat) was adopted.

\section{Results}\label{res}
\subsection{Spectra}
\begin{figure}
    \centering
    \includegraphics[width=\textwidth]{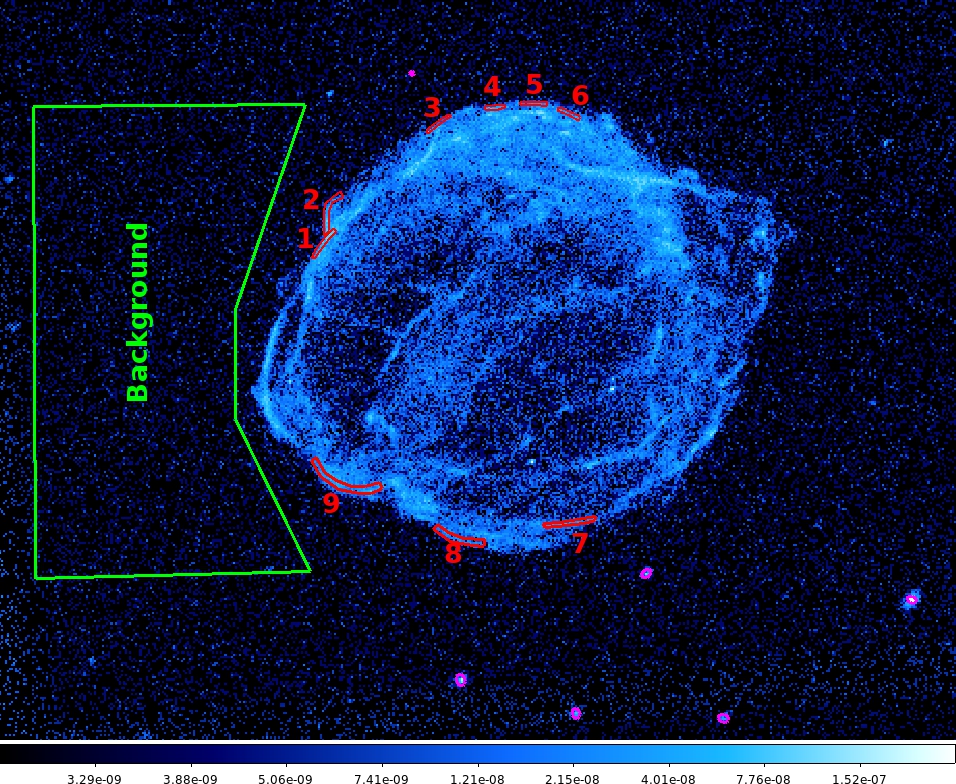}
    \caption{Chandra flux map of Kepler's SNR in $4.1-6$ keV band. Source regions are marked with red polygons, Background region is marked with the green polygon and magenta ellipses are the \cite{2022ApJ...926...84C} point sources used for the astrometric alignment.}
    \label{fig:reg}
\end{figure}
We selected 9 filamentary regions from north to south, which are clearly visible in the $4.1-6$ keV energy band, were the synchrotron radiation dominates the emission, and are all located on the rim of the shell. These regions are shown in red in Figure \ref{fig:reg}.
We extracted the spectra from the 2006 and 2014 observations, revising the location of the extraction regions accounting for the expected proper motion between the two epochs.
We analyzed the background by extracting the spectrum from the green region in Figure \ref{fig:reg}, and adopting a phenomenological model to fit it.
The source spectra were then fitted using the loss-limited model to describe the non-thermal feature of the spectra. 
To this model we added the background model (accounting for the different areas of the extraction regions), and a thermal component from optically thin plasma in non-equilibrium of Ionization (\texttt{vnei}) to describe the thermal features present in the spectra.
After obtaining the best fit model for each region, we measured the flux in the $4.1-6$ keV band.
In order to do so, we multiplied the previously defined model for the \texttt{cflux} model within XSPEC and fixed the parameters of the thermal component to their best-fit values.
The best-fit parameters are shown in Table \ref{tab:ecut}.
\begin{table}
    \centering
    \caption{Best-fit $\varepsilon_0$, Flux and velocity best-fit from regions labeled from 1 to 9. Errors for $\varepsilon_0$ and flux are at the 68\% confidence level. Errors for shock velocity are at 90\% confidence level.} 

\begin{tabular}{cccccc}

\hline\hline
Region \#&\multicolumn{2}{c}{$\varepsilon_0$ (keV)}&\multicolumn{2}{c}{Flux $4.1-6$ keV (Log$_{10}$ erg cm$^{-2}$ s$^{-1}$)}&V$_{sh}$ (Km s$^{-1}$)\\
 & 2006 & 2014 & 2006 & 2014 &\\
\hline 
1&$0.50_{-0.04}^{+0.04}$   &$0.9_{-0.2}^{+0.3}$   &$-13.742\pm0.014$&$-13.68\pm0.03$&$3570\pm100$ \\
2&$0.73_{-0.06}^{+0.07}$   &$0.80_{-0.16}^{+0.23}$&$-13.801\pm0.014$&$-13.81\pm0.03$&$4690\pm120$ \\
3&$0.45_{-0.05}^{+0.07}$   &$0.41_{-0.10}^{+0.18}$&$-14.36\pm0.03$  &$-14.33\pm0.07$&$3690\pm50$  \\
4&$0.34_{-0.03}^{+0.04}$   &$0.38_{-0.09}^{+0.14}$&$-14.17\pm0.02$  &$-14.21\pm0.05$&$1870\pm70$  \\
5&$0.35_{-0.04}^{+0.04}$   &$0.19_{-0.03}^{+0.05}$&$-14.19\pm0.02$  &$-14.42\pm0.06$&$1520\pm100$ \\
6&$0.218_{-0.017}^{+0.019}$&$0.23_{-0.04}^{+0.06}$&$-14.22\pm0.02$  &$-14.20\pm0.06$&$1590\pm60$  \\
7&$0.32_{-0.02}^{+0.03}$   &$0.51_{-0.10}^{+0.16}$&$-13.963\pm0.018$&$-13.97\pm0.04$&$4160\pm70$  \\
8&$1.07_{-0.10}^{+0.12}$   &$0.89_{-0.16}^{+0.24}$&$-13.499\pm0.011$&$-13.51\pm0.03$&$7690\pm70$  \\
9&$0.43_{-0.03}^{+0.04}$   &$0.36_{-0.05}^{+0.07}$&$-13.988\pm0.018$&$-13.96\pm0.04$&$6000\pm100$ \\
 \hline
\end{tabular}
\label{tab:ecut}
\end{table}

\subsection{Shock velocity measurement}
To have an estimate of the shock velocity, we measured the proper motion from all the nine regions from 2006 to 2014.
We mirrored \cite{2008ApJ...678L..35K}, where a deeper description of the algorithm adopted can be found. 
We extract the one dimensional radial counts profiles of each filament from both the 2006 and 2014 epochs.
The profiles were extracted using Chandra events file with 0.492" pixel.
Each profile was then remapped into a 40 times denser grid using a quadratic interpolation, in a similar fashion to \cite{2016ApJ...823L..32W}.
The square root of the counts was taken as statistical uncertainty.
Then, we shifted the 2014 profile relative to the 2006 profile, minimizing the value of $\chi^2$. 
For our aim, we only reported the statistical errors, which are the 90\% confidence limits.
Once obtained the best-fit value for the angular shift ($\theta$), one can easily derive the shock velocities for each region, using the relation $v_{sh}=d\cdot\theta/t$, where $d$ is the distance of the remnant from us, and $t$ is the time difference between the 2 epochs (8 yrs).
The shock velocities with their uncertainties are reported in Table \ref{tab:ecut}.

\section{Discussions and Conclusions} \label{con}
\subsection{Flux Variability}
From a preliminary analysis of the flux measurement between the 2006 and 2014, (listed in Table \ref{tab:ecut}) the examined regions, with the exception of region 5, do not show any significant change in flux. 
This suggests that these regions have remained within a loss-limited regime throughout the eight-year duration.
Conversely, the measurements of flux in region 5 indicate a significant reduction of synchrotron radiation. 
This fading phenomenon may be due to two reasons: 
either reduced acceleration efficiency (due to higher $\eta$ (not the case) or lower shock speed) or increased efficiency of radiative losses (e,g, due to interaction with a locally higher magnetic field). 
Consequently, the radiation losses would manifest as the noticeable decrease in flux we observed, over this 8 year baseline.
We plan to further investigate this issue in a forthcoming paper.

\subsection{Cut-off photon energy $\varepsilon_0$ vs. the shock speed $v_{sh}$}
\begin{figure}
    \centering
    \includegraphics[width=\textwidth]{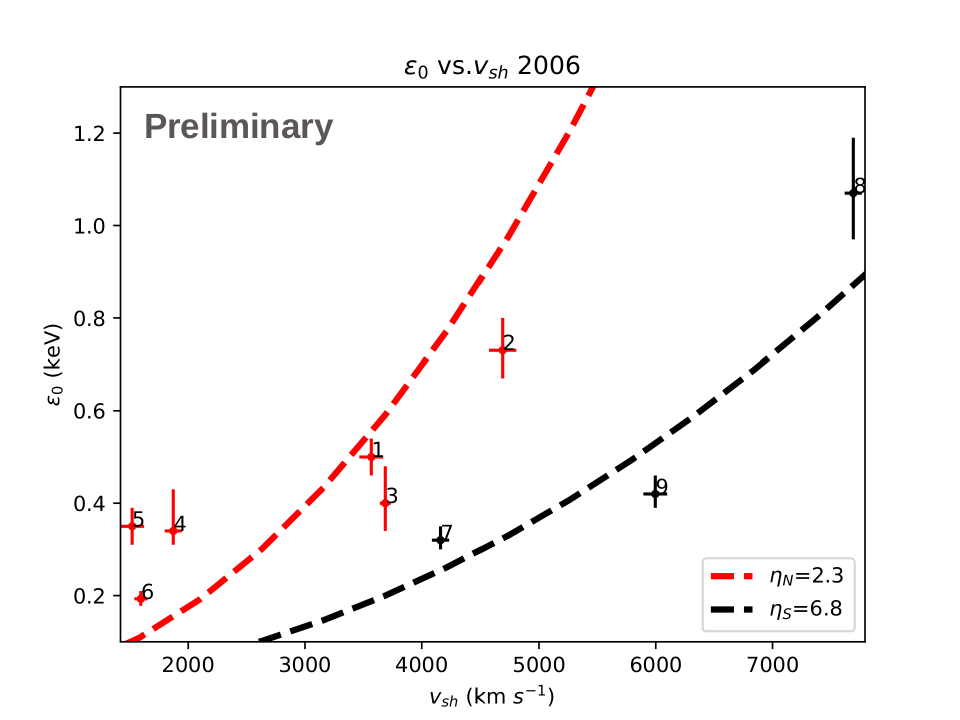}
    \caption{
    Synchrotron cutoff energy vs. current shock velocity for the year 2006.
    Red crosses mark northern regions (1-6) and the red dashed curve is the corresponding best-fit curve obtained from Equation \ref{eq:eovsvsh}. Black crosses mark southern regions (7-9) and the black dashed curve is the corresponding best-fit curve obtained from Equation \ref{eq:eovsvsh}.
    }
    \label{fig:ecut}
\end{figure}
By employing the same methodology outlined in our previous work \cite{2022ApJ...935..152S}, we present in Figure \ref{fig:ecut} the values of $\varepsilon_0$ (listed in Table \ref{tab:ecut}) obtained through the spectral fitting described in Section \ref{res}, plotted as a function of their corresponding shock velocity ($v_{sh}$) for the 2006 (the deepest observation analyzed).
We utilized different colors to distinguish between data points derived from southern regions (in black) and northern regions (in red).
The Figure \ref{fig:ecut} clearly illustrate the separation of the data points into two distinct clusters in either the epochs, representing the southern and northern regions. This observation corroborates the existence of two distinct regimes of electron acceleration within the same SNR, already identified by \cite{2022ApJ...935..152S}.
By fitting each of these two clusters using Equation \ref{eq:eovsvsh}, we can derive the corresponding best-fit values of the Bohm diffusing factor. For the southern regions, the retrieved Bohm factor is $6.8\pm 1.1$, as for the northern regions, the corresponding Bohm factor is $2.3\pm 0.4$.
The retrieved Bohm factors align well with the values obtained in our previous research, further strengthening the consistency and reliability of our findings.

\subsection{Conclusions}
We analyzed different \textit{Chandra} archival observations of Kepler's SNR in two different epochs: 2006 and 2014.
Our analysis has added some significant findings to our previous study \cite{2022ApJ...935..152S}, regarding particles acceleration and synchrotron emission dynamics in Kepler's SNR.
Firstly, our research has confirmed the existence of two distinct regimes of particle acceleration within Kepler's SNR. 
This strongly support the scenario in which the interaction between the shock and dense CSM in the northern regions results in the amplification of magnetic field turbulence.
Consequently, this interaction in the north leads to the acceleration of electrons in a loss-limited regime closer to the Bohm limit in respect to  the southern regions, where the shock, interacting with a tenuous medium, shows less efficient acceleration features.
Moreover, our study suggests a gradual decrease in synchrotron emission within the region 5.
Therefore we could have retrieved the first evidence of fading synchrotron emission from a specific region in the northern part of Kepler's SNR.
A more comprehensive analysis of the obtained results will be published in our upcoming paper.
\bibliographystyle{JHEP}
\bibliography{biblio}

\end{document}